\documentclass[%
 reprint,
 amsmath,amssymb,
 aps,
]{revtex4-2}

\usepackage{graphicx}% Include figure files
\usepackage{dcolumn}% Align table columns on decimal point
\usepackage{bm}% bold math
\usepackage{xcolor}
\usepackage[hidelinks,colorlinks=true,linkcolor=blue,urlcolor=blue,citecolor=blue,anchorcolor=blue]{hyperref}
\usepackage{gensymb}
\begin{document}

\preprint{APS/123-QED}

\title{Observation of Larmor-like precession of a single birefringent particle due to spin-dependent forces in tilted optical tweezers}% Force line breaks with \\

\author{Sauvik Roy}
\email{sr19rs022@iiserkol.ac.in}
\affiliation{Department of Physical Sciences, IISER-Kolkata, Mohanpur 741246, India}

\author{Nirmalya Ghosh}
\email{nghosh@iiserkol.ac.in}
\affiliation{Department of Physical Sciences, IISER-Kolkata, Mohanpur 741246, India}

\author{Ayan Banerjee}
\email{ayan@iiserkol.ac.in}
\affiliation{Department of Physical Sciences, IISER-Kolkata, Mohanpur 741246, India}

\author{Subhasish Dutta Gupta}
\email{sdghyderabad@gmail.com}
\affiliation{Department of Physical Sciences, IISER-Kolkata, Mohanpur 741246, India}
\affiliation{Tata Institute of Fundamental Research, Hyderabad, Telangana 500046, India}
\affiliation{Department of Physics, Indian Institute of Technology, Jodhpur 342030, India}

\date{\today}% It is always \today, today,

\begin{abstract}
We observe clear precessional motion of highly birefringent liquid crystal (LC) particles trapped in a spherically aberrated optical trap which is built around a tilted refractive index stratified medium. For input circularly polarized light, the breaking of azimuthal symmetry induced by the tilt leads to an asymmetric intensity distribution in the radial direction near the trap focal plane, which - in combination with the spin-orbit conversion effects for input circularly polarized light - results in non-uniform canonical and spin momentum densities in those regions. In addition, while the canonical momentum remains always oriented towards the axial direction, the spin momentum reverses direction along spatial loops in the radial direction. As a consequence, the total momentum precesses around the canonical momentum vector along elliptical spatial loops - akin to a Larmor-like precession of magnetic moment (total momentum in our case) around a magnetic field (canonical momentum). We probe this precession experimentally using the single trapped LC particles - with the direction of precession determined by the helicity of the input light and the precession frequency varying linearly with the laser power. Our experimental results are validated by numerical simulations of the system where we employ the Debye-Wolf theory for tight focusing in the presence of a tilted stratified media.   
\end{abstract}

\maketitle

%\tableofcontents

%\section{\label{sec:level1}Introduction} 
\textit{Introduction.-} Momentum is a crucial but curious protagonist in the realm of electromagnetic fields \cite{doi:10.1098/rstl.1884.0016}. Recently, the total momentum $(\boldsymbol{p})$ has been revealed to be composed of two components: canonical momentum $(\boldsymbol{p^o})$ \cite{bliokh2014extraordinary, berry2009optical} and spin momentum $(\boldsymbol{p^s})$, each with physically distinct origins and manifestations. This characterization has proved rather efficacious in elucidating fundamental photonic interactions in terms of the associated linear momentum, spin angular momentum (SAM) \cite{10.1119/1.14580, PhysRevA.48.656}, orbital angular momentum (OAM) \cite{PhysRevLett.88.053601, ALLEN200067}, and their mutual inter-conversions - commonly referred to as spin-orbit interaction (SOI) \cite{bliokh2015spin, doi:10.1126/science.aaa9519, PhysRevA.86.042103, brasselet2009prl, brasselet2009opticsletter, boyd2016jopt} of light. Typical manifestations of SOI can be seen in simple optical elements such as planar interfaces, lenses, and also in small particles \cite{Bliokh_2013,bliokh2011fsi} as well as complex structures e.g., photonic crystals, metamaterials, plasmonic structures \cite{zhang2020optica,tsesses2019nl,zhang2023pra}, etc. The in-plane and out-of-plane spatial Goos–Hänchen (GH) and Imbert–Fedorov (IF) \cite{Bliokh_2013} shifts respectively, of a paraxial beam undergoing reflection or refraction at planar interface and the spin Hall effects in scattering and focusing \cite{bliokh2015spin,Bliokh_2013} are well studied SOI related phenomena. Conversion from spin to intrinsic orbital angular momentum (IOAM) in an epsilon-near-zero (ENZ) \cite{PhysRevLett.118.104301} slab has been utilized in generating vortex beams out of a beam with no intrinsic orbital angular momentum (topological charge, $l=0$). A recent revelation of simultaneous spin-dependent directional guiding and wave-vector–dependent spin acquisition (spin-direction-spin coupling) in plasmonic crystals \cite{jeeban2023prl} facilitates the observation of the Hall effect and inverse Hall effect in the far field. 

Among these manifestations, certain events are recognized to be influenced by their inherent symmetry. The formation of spin-dependent vortices during the processes of focusing and scattering by small particles are found to result from the disruption of transverse translational symmetry and the violation of the electromagnetic duality \cite{PhysRevA.86.042103} in scattering, respectively. Breaking the cylindrical symmetry \cite{PhysRevA.82.063825} of a non-paraxial field is also seen to be accompanied by an enhancement of the spin Hall effect. In short, the SOI effects that govern the photonic response of an optical system are generally influenced by several factors encompassing both the field parameters and the material properties of the optical systems involved, and symmetry is one of the key ingredients in determining the nature of the interactions. 

In the context of SOI in optical tweezers \cite{aiello2015transverse, banzer2014pra, PhysRevA.105.023503, kumarprobing, singh2018pra, banzer2015tsamm}, beams with diverse rotationally symmetric intensity or polarization configurations are tightly focused to modify the field at the focus, and hence the magnitude or direction of the force experienced by micro-particles positioned in an axis-symmetric sample chamber with perfectly planar interfaces, can be altered. On the other hand, the fundamental Gaussian beam has been reported to exhibit helicity-dependent enhancement of transverse spin angular momentum (TSAM) \cite{roy2022pra} at the focal regions in the direction orthogonal to the symmetry-breaking axis when tightly focused through a tilted refractive index mismatched stratified medium. So, how the breaking of the axial symmetry by tilting the RI stratified medium affects the momentum distributions and subsequently modifies the forces acting on a microparticle is of paramount interest in the realm of micro-manipulation. In this paper, we demonstrate that a spin-responsive microparticle placed within a tilted sample chamber exhibits an unexpected concurrent precessional and partial orbital motion, the direction of the precession being determined by the helicity of the illuminating beam. The necessary insight into the underlying cause of the motion is provided by the spatial distributions of $\boldsymbol{p^o}$ and $\boldsymbol{p^s}$. Indeed, these distributions unveil a distinctive relationship between the two momentum components. While $\boldsymbol{p^o}$ primarily aligns along the beam propagation direction, the total momentum density $\boldsymbol{p}$ demonstrates a spatial precession, moving from one position to another, around the local $\boldsymbol{p^o}$ - mimicking the precession of magnetic moment around a magnetic field. Our work broadens the general understanding about the vectorial nature of canonical and the spin momentum density in an optical field, and can be potentially discernible in various other systems \cite{zemanak2018natcom} as well.   

\begin{figure}[h!]
\centering\includegraphics[width=1 \linewidth]{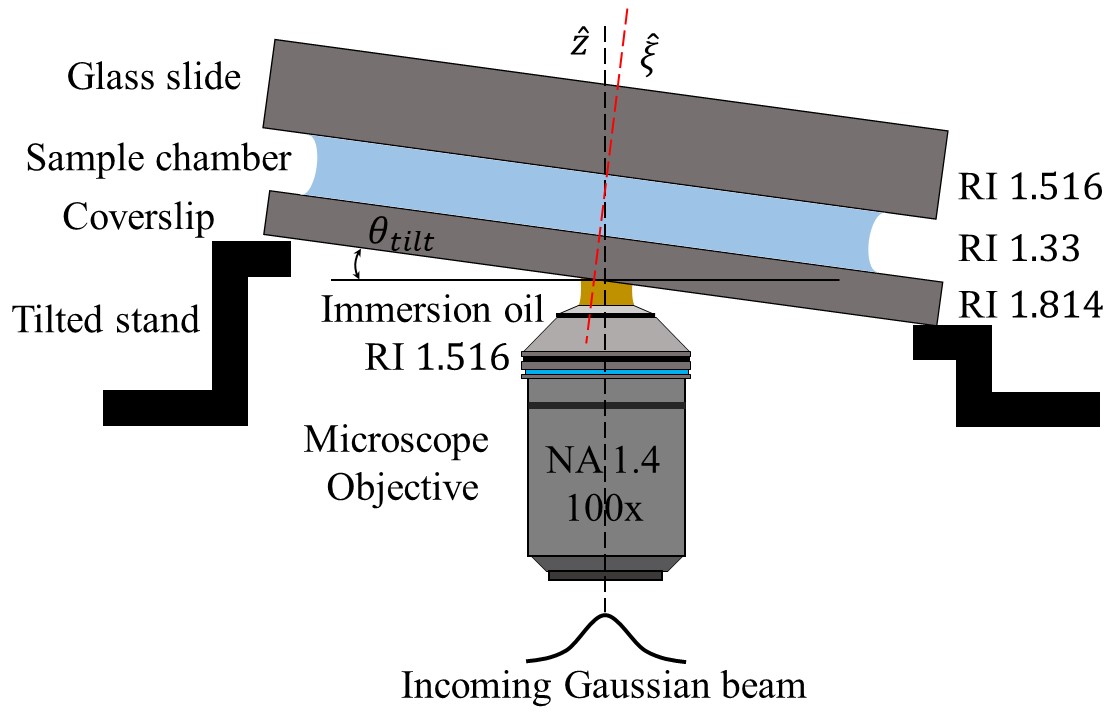}
\caption{Schematic of the Gaussian beam focusing through a tilted stratified medium.}
\label{fig:schem}
\end{figure}
\textit{Experiment.-} Our experimental configuration involves illuminating a four-layer stratified medium with refractive indices (RI) of $1.516$ (microscope objective immersion oil), $1.814$ (coverslip), $1.33$ (deionized water with $5$CB liquid crystal (LC) particles dispersed), and $1.516$ (top glass slide), respectively, using either a left (LCP) or right (RCP) circularly polarized Gaussian beam. Notably, the second layer, with a refractive index mismatch with the first, establishes an unconventional, stratified optical tweezers setup \cite{PhysRevA.87.043823, PhysRevA.85.033832, pal2020direct}. A tilt is applied to the coverslip by affixing a home-built mount below it as shown in Fig.~\ref{fig:schem}. The near-flat shape of the exit aperture of the objective lens and the necessity of focusing deep inside the sample chamber puts an experimental constraint on how much tilt can be applied, which comes out to a maximum possible value of $\sim 2.73^{\circ}$.

\begin{figure}[h!]
\centering\includegraphics[width=1 \linewidth]{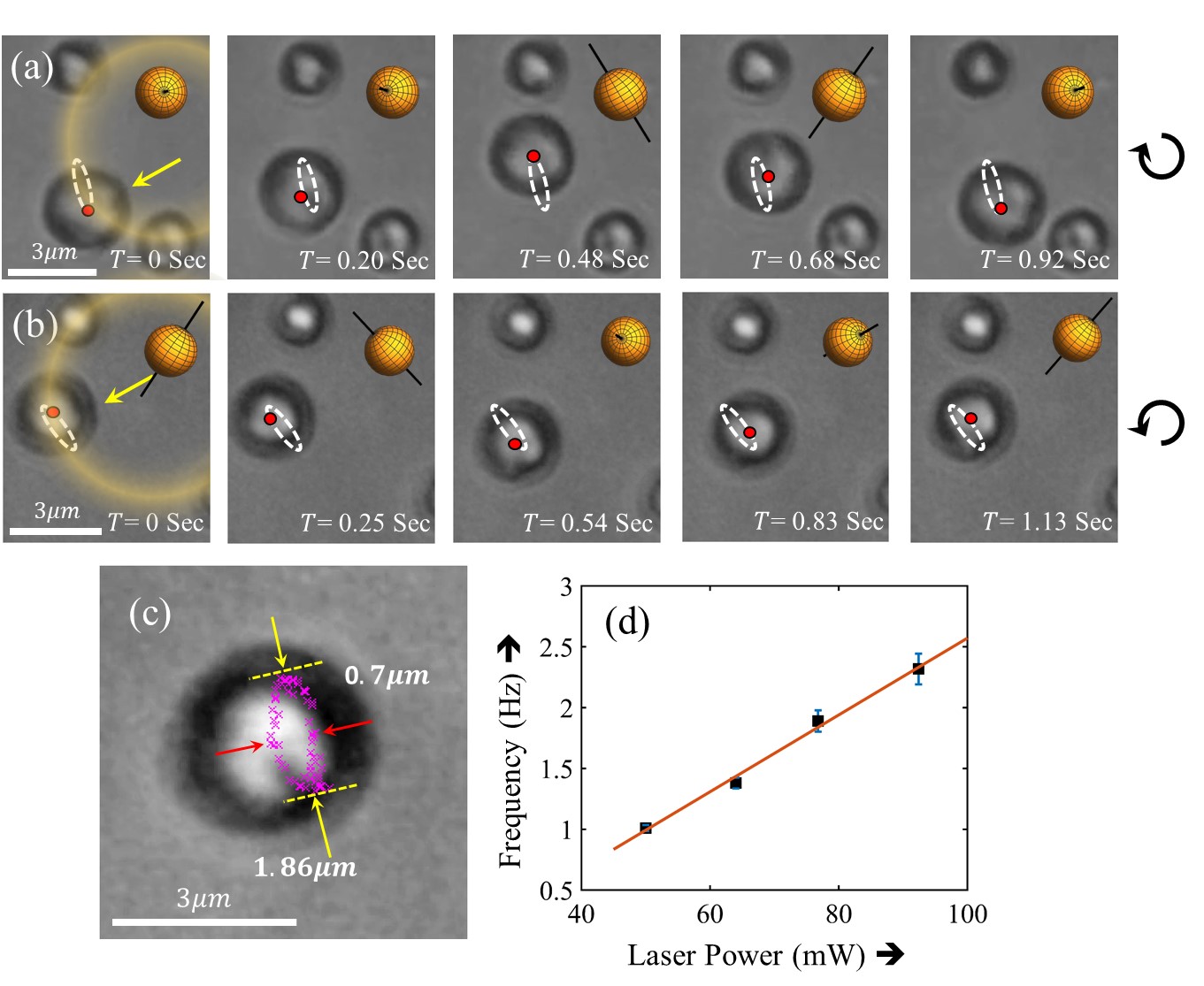}
\caption{ Concurrent precession and partial orbital movement of the designated particle in response to (a) a right circularly polarized (RCP) Gaussian beam and (b) a left circularly polarized (LCP) Gaussian beam. The semitransparent yellow ring delineates the off-axis high-intensity ring and the yellow spheres in each frame resemble the orientation of the liquid crystal particle. (c) A typical spatial trajectory of the particle when illuminated by a RCP Gaussian beam. (d) Linear variation of the precessional frequency with the incident laser power.}
\label{fig:experimental}
\end{figure}

The substantial refractive index (RI) contrast within the stratified medium leads to increased spherical aberration of the focused light. Consequently, uniform concentric high-intensity rings form, facilitating off-axial trapping of microparticles \cite{PhysRevA.87.043823, PhysRevA.85.033832, pal2020direct, PhysRevA.105.023503}. When a tilt is introduced, the regularity of the rings is disrupted, causing the intensity maxima to shift to one side of the ring, that is decided by the applied tilt axis. Consequently, particles are predisposed to being exclusively trapped on one side of the ring. This phenomenon is evident in Figs.~\ref{fig:experimental}(a), (b), where three LC particles are observed to be exclusively trapped on the left side of the ring. Note that any attempt to position particles on the right side of the ring by laterally moving the sample chamber proves unsuccessful.

The effect of such a distinctive trapping configuration on the dynamics of the trapped LC particles is the principal objective of our study. Indeed, single LC liquid crystal particles characterized by significant birefringence, are found to exhibit clear precession around an imaginary body axis, accompanied by partial orbital motion.  Thus, they actually appear to experience a spin force that we believe originates from the spin momentum and the corresponding spin angular momentum components, with the direction of precession and orbital motion being determined by the helicity of the input light (LCP/RCP beam). No such motion is observed for incident linear polarization. The motion for input LCP/RCP is substantiated by visually tracking the Maltese-like cross adhered to the particle, which appears by the imaging of the particles under cross-polarization, and is a manifestation of the orientation of the internal directors of the liquid crystal molecules. Figs.~\ref{fig:experimental}(a), (b) displays time-lapsed images of this complex motion for input LCP and RCP, with the orbital trajectory shown by the white dashed line. The corresponding videos (Video 1 and 2, for LCP and RCP, respectively) are provided in the Online Supplementary Information. A typical orbital trajectory appears to be elliptical measuring $1.9 \mu m~\times 0.7\mu m$ in the major and minor axes, respectively, for input RCP (Fig.~\ref{fig:experimental}(c)) and $1.6 \mu m~\times 0.7 \mu m$  for input LCP. The precession rate of the particle appears to be influenced by the power of the beam, increasing with higher laser power, as shown in Fig.~\ref{fig:experimental}(d).

%\section{THEORETICAL ANALYSIS}

\textit{Theoretical formalism.} We invoke the pioneering research conducted by Debye and Wolf \cite{richards1959electromagnetic, wolf1959electromagnetic} for tight focusing of a well-collimated beam to understand the aforementioned complex motion. In this formalism, the refracted spherical wave from an aplanatic lens is portrayed as a vectorial superposition of plane waves with infinite spatial harmonics. The intricate transformations of the complex electromagnetic field within the high numerical aperture (NA) lens are effectively elucidated by the lens's transfer function. This transfer function typically incorporates Fresnel's reflection and transmission coefficients \cite{PhysRevA.85.033832, torok1997stratified}, providing a more realistic representation of experimental configurations with multiple parallel interfaces perpendicular to the lens axis, delineating different media within the lens's image space. The efficacy of the transfer function mechanism is evident when the stratification axis aligns with the lens axis. In this scenario, despite multiple scattering events within the stratified medium, ray optics dictates that all constituent plane waves consistently stay within their respective meridional planes. However, deliberate tilting of the stratified medium relative to the lens axis introduces a change in the plane of incidence at the initial interface \cite{roy2022pra}. This alteration in the plane of incidence, affecting the subsequent superposition of constituent plane waves, leads to a redistribution of electromagnetic fields in the regions beyond the lens.

Due to the non-paraxial nature of the problem, introducing the tilt-dependent change in the angle of incidence into a new transfer function proves to be non-trivial. Notably, as this formalism essentially involves a vectorial sum of complex fields, the problem can be conveniently decoupled into two distinct components: focusing and the independent propagation of plane waves through the stratified medium \cite{munro2018tool}. This separation allows for the continuous tracking of the spin or polarization evolution during the propagation of light through the medium. Furthermore, it opens avenues for leveraging the full potential of modern computational advancements. According to this formalism, the time-independent monochromatic electric field at a point $p$ $(\boldsymbol{r}_{p})$ in a homogeneous medium of RI $n_0$ is
\begin{equation}
        {\bf E}(\boldsymbol{r}_{p})=-i\beta\iint_{\Omega} \frac{{\boldsymbol{\epsilon} }(s_{x},s_{y} )}{s_{z}} e^{in_{0}k_{0}({\bf s}.\boldsymbol{r}_{p})} \,ds_x\,ds_y
\end{equation}
Here, $\beta=\frac{n_{0} k_{0} f}{2 \pi}$, ${\boldsymbol{\epsilon}}(s_{x},s_{y})$ is the electric field spectra on the reference Gaussian sphere, ${\bf s}=(s_{x},s_{y},s_{z})$ is typically the direction of an incoming ray in the geometric description, $k_{0}$ is the free space wave number, $\Omega$ is the solid angle formed by the geometric rays converging from the lens. For brevity, only the transformation of the electric field components is discussed in this paper. The temporal factor is assumed to have the form $\text{exp}(-i \omega t)$.

As mentioned earlier, the transfer function that connects the incident amplitude $\boldsymbol{\epsilon}^{in}$ to the refracted field amplitude $\boldsymbol{\epsilon}(s_{x},s_{y})$ involves a systematic arrangement of three SO($3$)  rotation matrices which corresponds to the decomposition of the incident field into TE and TM components via a coordinate system rotation. The explicit form of the transfer function is  
\begin{equation}
    A=R_{z}(-\phi)R_{y}(\theta)R_{z}(\phi).
\end{equation}
Where, $R_{z}(\phi)$ and $R_{y}(\theta)$ are rotation matrices along $z$ and $y$ axis respectively. Following the Jones vector, the resultant field amplitudes on the reference sphere read as
\begin{equation}
    {\boldsymbol{\epsilon}}(s_{x},s_{y})=\sqrt{\cos{\theta}}A{\boldsymbol{\epsilon}}^{in}.
\end{equation}
%\begin{widetext}
\begin{equation}
{\boldsymbol{\epsilon} }(s_{x},s_{y})
=
\begin{bmatrix}
a-b\cos{2\phi} & -b\sin{2\phi} \\
-b\sin{2\phi} & a+b\cos{2\phi} \\
-c\cos{\phi} & -c \sin{\phi}
\end{bmatrix}
\begin{bmatrix}
{\epsilon}^{in}_{x}\\{\epsilon}^{in}_{y},
\end{bmatrix}
\end{equation}
%\end{widetext}
where, $\boldsymbol{\epsilon}^{in}= [{\epsilon}^{in}_{x},{\epsilon}^{in}_{y}]^T$  corresponds to the Jones vector of the incident polarization state. 
\begin{figure*}[t]
\centering\includegraphics[width=1 \linewidth]{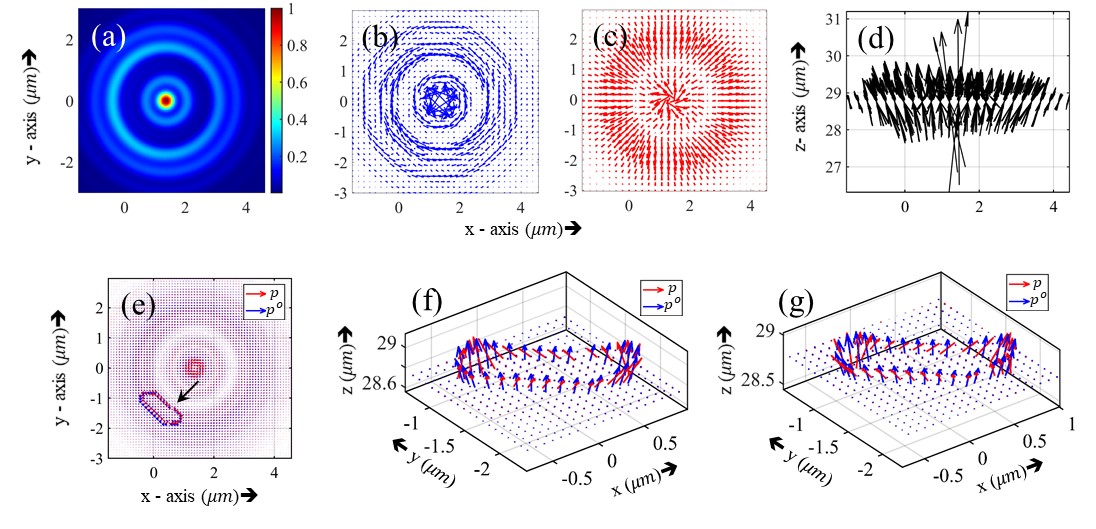}
\caption{(a) Irregularity in the simulated off-axial intensity ring at a transverse plane very close to the last interface ($z=28.75\mu m$). The stratified medium is tilted by an angle of $2.73^{\circ} $. (b) and (c) are the 2D vector plots of the transverse ($x$ and $y$) components of Belinfante's spin momentum $(\boldsymbol{p^{s}})$ and canonical momentum $(\boldsymbol{p^{o}})$ respectively for the incident RCP beam. (d) The helicity independent $\boldsymbol{p^{o}}$ can be seen predominantly oriented along the beam propagation direction with minimal transverse components due to tilting and divergence. (e) One proposed closed trajectory of nearly elliptical shape is highlighted in the XY view of the 3D vector plot of $\boldsymbol{p^{o}}$ and total momentum $(\boldsymbol{p})$. (f) and (g) contain the 3D orientations of the $\boldsymbol{p^{o}}$ and total momentum $(\boldsymbol{p})$ for the LCP and RCP illuminations respectively. The vectors along the proposed trajectory in (e) are scaled up by a factor of 15 for enhanced visualization of the point-to-point precession of $\boldsymbol{p}$ around $\boldsymbol{p^{o}}$.}
\label{fig:_po_ps}
\end{figure*}
The effect of the stratified medium can be incorporated through an independent decomposition of each plane wave coming from the lens into TE and TM modes \cite{munro2018tool} with the help of another transfer matrix for the stratified medium at the observation plane transverse to the stratification axis $\boldsymbol{\hat{\xi}}$. Owing to the linearity of the system, the total electric field at a point $(x,y,z)$ can be found as a superposition of forward and backward propagating fields
\begin{multline}
	{\bf E}^{\pm}(x,y,z)=-i\beta\iint_{\Omega} [a^{\pm}_{TE}(z) {\bf u}^{\pm}_{TE} + a^{\pm}_{TM}(z) {\bf u}^{\pm}_{TM}]\\ e^{in_{0}k_{0}(s_{x}x+s_{y}y)} \,ds_x\,ds_y
\end{multline}
Here, $+$ and $-$ signs are used to denote forward and backward propagating waves respectively, ${\bf{a}}$'s are the amplitudes, and $\bf{u}$'s are the unit vectors along TE and TM components.   

By virtue of the formulation of the Debye-Wolf integral, the electric field is obtained in the laboratory frame, say $L_{1}$, whose origin coincides with the geometric focal point $O_{1}$ of the lens. The transverse axes are defined by the polarization of the incident beam. But in the case of a tilted stratified medium, instead of solving the integral in $L_{1}$ frame, it is convenient to solve in a frame $L_{2}$ which is rotated by an angle $\theta_{tilt}$ and whose origin coincides with the focal point $O_{1}$. At this step, all the vector quantities $\bf{E} $, $\boldsymbol{s}$, and $\boldsymbol{r}_{p}(x,y,z)$ are transformed by the $3\times3$ rotation matrix $\bf{T}$ which takes care of both the axis and the tilting angle $\theta_{tilt}$. In the rotated frame, the incident field strength takes the form $\textbf{E}^{'}_{L_{2}}=\textbf{T}\textbf{E}$. Next, the matrix formalism gives us the required field $\textbf{E}^{'}$ at the required plane in the rotated frame. The electric field in the laboratory frame $L_{1}$ is obtained by the inverse transformation $\textbf{E}_{final}=\textbf{T}^{-1}\textbf{E}^{'}$. The rotation matrix $\textbf{T}$ and its inverse $\textbf{T}^{-1}$ forms the similarity transformation mechanism that facilitates the modelling of the tilted stratified medium via coordinate rotation and consequently simplifies the process of finding the field components in the laboratory frame $L_{1}$. 

Having obtained the electric and magnetic fields in a nonmagnetic medium (i.e., $\mu = \mu_{0}$) with refractive index $n$, the canonical $(\boldsymbol{p^{o}})$ and spin momentum $(\boldsymbol{p^{s}})$ densities are found according to the explicit forms\cite{bliokh2014extraordinary, sahareviewtsam}:
\begin{gather}
    \boldsymbol{p^{o}}=\frac{1}{4\omega} \text{Im}[\varepsilon_{0} n^2 \textbf{E}^{*}.(\boldsymbol{\nabla}) \textbf{E} + \mu_{0} \textbf{H}^{*}. (\boldsymbol{\nabla}) \textbf{H}]
\\
    \boldsymbol{p^{s}}=\frac{1}{2} \boldsymbol{\nabla} \times \textbf{s}
\end{gather}
%\begin{equation}
%    \boldsymbol{p^{o}}=\frac{1}{4\omega} Im[\epsilon_{0} \textbf{E}^{*}.(\nabla) \textbf{E} + \mu_{0} %\textbf{H}^{*}. (\nabla) \textbf{H}]
%\end{equation}
%\begin{equation}
%    \boldsymbol{p^{s}}=\frac{1}{2} \nabla \times \textbf{s}
%\end{equation}
Where, the spin angular momentum (SAM) density $\textbf{s}$ is given by:
\begin{equation}
    \textbf{s}=\frac{1}{4\omega} \text{Im}(\varepsilon_{0} n^2\textbf{E}^{*} \times \textbf{E} + \mu_{0} \textbf{H}^{*} \times \textbf{H})  
\end{equation}
where, $\varepsilon_{0}$ and $\mu_{0}$ are the permittivity and permeability of free space. The intrinsic dispersion of the medium has been taken care of by the permittivity $(\epsilon = \varepsilon_{0} n^2)$ and permeability $(\mu)$ of the medium.

%\section{\label{sec:level1}Results and Discussion}

%\subsection{Analysis of the precession:}

\textit{Numerical simulations.} We now attempt to understand this complex rotational trajectory by simulating our experimental system. We thus develop a MATLAB code that incorporates the above-mentioned Debye-Wolf method \cite{roy2022pra} with parameters from the real experimental setup. The interface positions are set at $-111\mu m$, $19\mu m$, and $29\mu m$ from the geometric focus of the lens. The same experimental tilt angle of $2.73^{\circ}$ is set in the simulation. Because of the tilt, a nonuniform distribution of the intensity across the rings accompanying an overall focal shift can be observed in the simulated intensity pattern at a transverse plane close to the last interface (Fig. \ref{fig:_po_ps}(a)). To visually depict the motion, two 2D quiver plots showcasing the transverse components ($x$ and $y$) of the spin momentum ($\boldsymbol{p^{s}}$) also called the Belinfante's spin momentum (BSM) and canonical momentum $\boldsymbol{p^{o}}$ are generated (Figs. \ref{fig:_po_ps}(b), (c)). Unfortunately, these plots do not provide a clear understanding of the motion. Therefore, a 3D quiver plot of canonical momentum $\boldsymbol{p^{o}}$ (Fig. \ref{fig:_po_ps}(d)) is generated subsequently, revealing that $\boldsymbol{p^{o}}$ is predominantly oriented along the stratification axis $\boldsymbol{\hat{\xi}}$ with minimal $x$ and $y$ components along the transverse directions. However, a distinct distribution characterizes the spin momentum $\boldsymbol{p^{s}}$, where its $z$-component undergoes no change in sign on the two sides of a high-intensity ring, but the $x$ and $y$ components exhibit helicity dependence, as depicted in Fig. \ref{fig:_po_ps}(b). Clearly, the understanding of the reason for the precessional motion is contingent upon the nature of the distribution of $\boldsymbol{p^{o}}$ and $\boldsymbol{p^{s}}$ - a matter that will become apparent subsequently.

Due to the elliptical nature of the trajectory and the input power dependency of the precession rate - in the first approximation - the rate of change of total momentum can be equated with $\boldsymbol{p} \times \boldsymbol{p^o}$ through a relation of the form $\frac{d\boldsymbol{p}}{dt} \propto \boldsymbol{p}\times \boldsymbol{p^o}$. Quite interestingly, this is nothing but the well-known Bloch equation. The canonical momentum $\boldsymbol{p^o}$ plays the role of the magnetic field and the total momentum $\boldsymbol{p}$ is the vector that precesses around $\boldsymbol{p^o}$. The correspondence between $\boldsymbol{p^o}$ and the external magnetic field is further emphasized by the fact that the spin momentum $\boldsymbol{p^s}$ represents a solenoidal current $\boldsymbol{\nabla}.\boldsymbol{p^s}=0, \int \boldsymbol{p^s} \,d\text{v} = 0 $ and does not contribute to energy transport. Consequently, the linear correlation between the precession frequency and the external magnetic field, along with the linear relationship between precession frequency and laser power or $\boldsymbol{p^o}$, substantiates the soundness of our analogy. It is to be noted that the transverse Belinfante's spin momentum crucially contributes to the precession of the total $\boldsymbol{p}$ around $\boldsymbol{p^o}$. The point-to-point variation of $\boldsymbol{p}$ and $\boldsymbol{p^o}$ for both LCP and RCP along one possible trajectory of the particle is shown in Figs. \ref{fig:_po_ps}(e), (f), (g). As the particle moves point by point, it is influenced by the local $\boldsymbol{p}$ which dictates the closed path, and the Maltese-like cross on the particle can be seen precessing due to the total momentum $\boldsymbol{p}$ while remaining trapped in the same ring. The proportionality constant $\alpha$ resembles the gyromagnetic ratio of the trapped lc particle. 

For an intuitive grasp of the extent of the trajectory, we present the 3D quivers of $\boldsymbol{p^o}$ and $\boldsymbol{p^s}$ in Figs. \ref{fig:_po_ps}(e), (f), (g). It is readily discernible that, in a region indicated by the black arrow in Fig. \ref{fig:_po_ps}(e), the total momentum density $\boldsymbol{p}$ undergoes spatial precession around $\boldsymbol{p^o}$. Notably, the blue arrows denoting $\boldsymbol{p^o}$ and the red arrows denoting $\boldsymbol{p}$ along the trajectory are magnified by a factor of $15$ for enhanced visualization. The simulated trajectory's dimensions are measured to be $\sim 0.56\mu m$ in length and $\sim 1.67\mu m$ in breadth, aligning well with the experimentally observed dimensions (length $\sim 0.7\mu m$ and breadth $\sim 1.9\mu m$). In addition, the position of the trajectory in the simulated plot and the experimentally observed trajectory align with each other. The point-to-point variation of $\boldsymbol{p}$ around $\boldsymbol{p^{o}}$, the trajectory extent, and the input power dependency of the precession rate collectively demonstrate the precession of $\boldsymbol{p}$ around $\boldsymbol{p^o}$ in a tightly focused tilted stratified medium in optical tweezers.

\textit{Conclusion.-} In conclusion, we observe an intriguing precessional motion of a highly birefringent particle in optical tweezers developed around a tilted stratified medium. Due to the breaking of the azimuthal symmetry in the problem, the spin momentum becomes radially inhomogeneous. This, coupled with the fact that it reverses direction along spatial lobes in the radial direction while the canonical momentum is always aligned axially, leads to a resultant precession of the total momentum about the canonical momentum. The direction of precession is controlled by the helicity of the input light beam, with the rate depending on the laser power. Indeed, the association of canonical momentum with the magnetic field and the helicity-dependent total momentum with magnetic moment leads to an optical analogy of a spinning particle in an external uniform magnetic field, represented by the ubiquitous Bloch equation. Here, however, the equivalents of both the magnetic moment and the external magnetic field are integral components of light itself, thus revealing another exotic manner in which it manifests itself in the mesoscopic world. Indeed, the use of more complex input beam profiles in such tilted optical tweezers could yield even more interesting results, and generate diverse and unexpected avenues in particle manipulation.

\begin{acknowledgments}
Sauvik Roy is thankful to the Department of Science and Technology (DST), Government of India for the INSPIRE fellowship.

\end{acknowledgments}

\appendix

\nocite{*}

%\bibliographystyle{apsrev4-2}

\providecommand{\noopsort}[1]{}\providecommand{\singleletter}[1]{#1}%

\end{document}